\documentstyle[aps,preprint]{revtex}
\begin{document}

\centerline{\bf Analysis of Ly ${\alpha}$ absorption lines in the
vicinity of QSOs}
\vskip 0.1 true in
\centerline{\bf R. Srianand}
\centerline{\it Inter University centre for Astronomy and Astrophysics}
\centerline{\it Post bag 4, Ganeshkhind, Pune 411007, India}
\centerline{and}
\centerline{\bf Pushpa Khare}
\centerline{\it Department of Physics, Utkal University,
Bhubaneswar-751004, India}
\centerline{email: anand@iucaa.ernet.in and pk@utkal.ernet.in}
\vskip 0.5 true in

\begin{abstract}
We have compiled, from the literature, a sample of Ly ${\alpha}$ forest
lines in the spectra of 69 QSOs, all observed with a resolution between
60 to 100 km s$^{-1}$. The sample is studied for proximity effect. We
have tried to account for the effect of blending which is inherent in
the intermediate resolution sample, by calculating the column density
distribution, using an effective velocity dispersion parameter, from
the observed equivalent width distribution. The use of this column
density distribution in the proximity effect analysis reduces the
background intensity values by a factor of 2 to 3 compared to the
values obtained by using the column density distribution obtained from
high resolution observations. Evidence is presented for a weak
correlation between the effective velocity dispersion parameter and
equivalent width. Such a correlation if present can increase the
background values by a factor of up to 1.5. Considerations of proximity
in the spectra of 16 QSOs, from our sample, exhibiting damped Ly
${\alpha}$ lines gives a background intensity which is 3 times smaller
than the values obtained from the whole sample, confirming the presence
of dust in the damped Ly $\alpha$ systems. Lines close to the QSOs are
shown to be marginally stronger and broader compared to lines away from
the QSOs.  The Ly $\alpha$ lines with absorption redshift larger than
emission redshift are shown to be uncorrelated with QSO luminosity,
radio loudness or optical spectral index. These lines occur more
frequently at high redshifts. Their presence is correlated weakly with
the presence of associated metal line systems. The possibilities that
the QSO emission redshift is considerably higher and that either the Ly
$\alpha$ clouds or the QSOs have peculiar velocities are considered. It
is argued that a combination of both these phenomenon may be required
to account for the presence of these lines.

\end{abstract}

\centerline{\bf 1. INTRODUCTION}

An inverse trend in the distribution of Ly $\alpha$ forest lines
near the QSOs was first reported by Carswell et al. (1988) and
was later confirmed by various groups. Using non parametric
Q-Test Murdoch et al. (1986; here after MHPB), and Lu et
al.(1991) showed that fewer lines, than the number predicted by
the best fit power law redshift distribution, are present in the
regions close to the QSOs. This was interpreted as being due to
the increased ionization of Ly ${\alpha}$ clouds near the QSOs
due to their radiation field, and was used, first by Bajtlik et
al.(1988; here after BDO), to obtain the intensity of background
UV radiation field.

Recently Bechtold (1994) with an extended sample of intermediate
resolution, FWHM${\simeq 1 \AA}$, spectra showed that proximity
effect is present near QSOs at very high significance level.
The background intensity values obtained by her are consistent
with earlier results, and are higher than the values expected
from the QSOs and other AGNs.

Intermediate resolution samples of Ly ${\alpha}$ lines consist
of large number of QSO sight lines and sample the redshift space
more or less uniformly. These are thus free from the effects
induced by peculiar  distribution of lines along one or two QSO
sight lines.  However due to high number density of Ly
${\alpha}$ lines, the lines observed with intermediate resolution
are usually blends and may lead us to wrong
interpretations. In this paper we try to study the effect of
various sources of uncertainties in the background intensity
calculations based on proximity effect.  For this purpose we
have collected intermediate resolution spectra of large number
of QSOs, with similar S/N, from the literature.  The details of
the data used are given in section 2. In section 3 we study the
general properties of lines with ${\rm z_{abs}<z_{em}}$. In
section 4, we discuss the proximity effect and the effect of
curve of growth and the low resolution on the calculations of
background intensity. The properties of lines close to the
emission redshift are discussed in section 5.  In section 6 we
discuss the Ly ${\alpha}$ lines with ${\rm z_{abs}>z_{em}}$.  We
consider two possible origins for these lines, and for each of
these hypothesis we calculate the background intensity.
Conclusions are presented in section 7.

\centerline{\bf 2. DATA SAMPLE}

Useful information regarding the QSOs used in our sample along
with the references are given in Table 1.  Emission redshift of
the QSOs quoted in the literature are given as ${\rm z_{em}}$.
However it is known that most of these emission line redshifts
are obtained from high ionization lines and may not represent
the systemic redshifts.  Tytler and Fan (1992)  have given
formulae to obtain the systemic redshifts. We have used these,
with all available emission lines, to obtain the average values
of corrected redshifts, ${\rm {z^c_{em}}}$ which are also given in
Table 1.  ${\rm z_{min}}$ is the larger of the observed minimum
and the redshift corresponding to the Ly $\beta$ emission. ${\rm
z_{max}}$ is the maximum observable redshift for Ly $\alpha$
lines. Since in this paper we also study the Ly ${\rm \alpha}$
lines with ${\rm z_{abs}>{z^c_{em}}}$ Ly ${\alpha}$ lines,
hereafter NVLs(negative velocity lines), the maximum limit is
taken as ${\rm z_{max}}$ rather than ${\rm {z^c_{em}}}$.  E is the
minimum detectable equivalent width of a line at 5 $\sigma$
level in each spectra. In most of the cases these  values are
taken as given by of the authors of the parent references.
Wherever these are not given we have taken E to be five times
the largest observable error in the equivalent width of any line
in the relevant redshift range. ${\rm f_\nu}$ is the QSO
continuum flux at the Lyman limit. The values are calculated by
extrapolating the continuum flux at the rest wavelength of
$\lambda (1450)$ to the Lyman limit. The spectral index $\alpha$
is also given in the Table. For QSOs for which the value of
$\alpha$ is not available we use $\alpha\;=\;0.66$. RL is the
radio loudness, defined as

$${\rm RL = {log\bigg({S(5GHz)\over S(\lambda 1450)}\bigg)}.}$$

Wherever 5 GHz radio measurements are not available the radio flux
was extrapolated from the other available radio measurements,
assuming a powerlaw with 0.3 as the spectral index. Note that
our sample has a large number of QSOs common with the sample
compiled by Bechtold (1994). Out of 67 QSOs used here, 54 are
from her sample. All the known metal lines and the Ly ${\alpha}$
belonging to metal line systems are considered as excluded
regions. The Ly ${\alpha}$ lines with rest equivalent width
$>5\AA$ are also taken to be excluded regions as they may well
be damped Ly ${\alpha}$ candidates.

\centerline{\bf 3. REDSHIFT AND EQUIVALENT WIDTH DISTRIBUTION
OF Ly ${\alpha}$ CLOUDS}

In order to calculate the expected number of Ly ${\alpha}$ lines
within the region near a QSO one should know the evolutionary
properties of the lines with ${\rm z_{abs} < z_{em}}$ without any
ambiguity. Here we use our extended data to determine these
properties. It is customary to describe the redshift and
equivalent width distribution of Ly $\alpha$ lines as,

\begin{equation}
{\rm{\partial^2N\over \partial z\partial W}\;} = {\rm
A(1+z)^\gamma
\exp{-W\over W_*}},
\end{equation}

where ${\rm{\partial^2N\over \partial z\partial W}\;}$  is the
number of lines per unit redshift interval, per unit equivalent
width, per line of sight.  We used the maximum likelihood method
described by MHPB, to calculate ${\rm W_*}$ and ${\gamma}$
assuming them to be constant, independent of z and equivalent
width, using the data described in the previous section. In
order to have an unbiased sampling in the redshift space we
confine ourselves to the redshift range between 1.7 to 3.7 and
consider only those lines which are more than 8 Mpc (for ${\rm
q_o = 0.5}$) away from the QSO.  The calculated values of
${\gamma}$ for various rest equivalent width cutoffs, ${\rm
{W_r}^{min}}$, are given in table 2. Also given in table 2.
are the values of KS probability that the largest difference
between the observed redshift distribution and the distribution
described by eqn (1), with the value of ${\gamma}$ given by the
fit, occurs by chance.

A glance at Table 2 reveals that there is a clear trend of increase in
${\gamma}$ with an increase in ${\rm {w_r}^{\min}}$, consistent with
Bechtold's result which is expected as {80\%} of the QSOs are in common
with her sample. As noted by her the ${\gamma}$ values are lower than
the previous estimates made with intermediate resolution observations.
These are also somewhat lower than the estimates of ${\gamma}$ made
with high resolution sample (Srianand and Khare 1994; here after Paper
I).  The significance of the fit is very low for ${\rm {W_r}^{min}} \le
0.2\AA$ but is acceptable for higher values of ${\rm {W_r}^{min}}$.
This is primarily because, though the S/N of the spectra are good
enough to detect even very weak lines, the poor resolution prevents us
to deblend weak lines from the blended features. Low resolution thus
introduces an incompleteness in the sample and also introduces a bias
towards detecting weak lines more frequently at the lower redshifts, as
the number density there is less than that at the higher redshift. This
gives a smaller $\gamma$ value, as well as produces  an artificial
differential evolution as was noticed by Liu {\&} Jones (1988). The
dependence of ${\gamma}$ on ${\rm {W_r}^{min}}$ is however weaker, for
${\rm {W_r}^{min}}\ge 0.25\AA$, than that found by Acharya and Khare
(1993).Press, Rybicki and Schneider(1993), using a novel method, have
obtained the value of $\gamma$ to be 2.46$\pm$ 0.37. Their method makes
use of the depression in the continuum on the short wavelength side of
the Lyman alpha emission line and is free from any bias introduced due
to the resolution used. Their data set is different and smaller
compared to our data set. The difference between their value and the
value obtained above is, however, not the effect of resolution as even
the high resolution data (paper I) gives values smaller than their
value.

The values of the equivalent width distribution parameter ${\rm
W_*}$, calculated using maximum likelihood method, for various
${\rm {W_r}^{min}}$ are also given in table 2. It is clear that
unlike ${\gamma}$, ${\rm W_*}$ is independent of the rest
equivalent width cutoff. As expected for the blended
distribution, the values of ${\rm W_*}$ obtained here are higher
than the corresponding values obtained from high resolution
observations (Paper I).

We also performed two other statistical tests in order to
estimate the goodness of fits. We used chi-square test to
determine the $\chi^2$ per degrees of freedom between the
calculated and expected number of lines based on equation (1),
using values of $\gamma$ given by the fit, for individual QSOs.
These values are also given in table 2.  The $\chi^2$ values
indicate that the observed number of lines are not markedly
different from the expected number in individual QSOs.  Also
given in the table are the values of weighted average ${\rm
<Q-0.5>}$, as described by MHPB. Note that our values are nearly
equal to zero, as one expects in the case where the assumed
global distribution is consistent with the observed distribution
in individual QSOs.  In order to confirm this we performed
Wilcoxon test described by Lu et al. (1991). Our results
indicate that the probability that the observed values of ${\rm
<Q-0.5>}$ are distributed normally around zero, as is expected
for a good fit, is 0.95 for ${\rm {W_r}^{min}=0.3\AA}$. We
define a parameter Y,  = ${\rm n_e-n_o}$, where ${\rm
n_e\;and\;n_o}$ are the expected and observed number of lines in
individual QSOs. If the distribution along individual sight
lines is well described by the determined values of ${\gamma}$
one expects Y to distribute normally around zero. The calculated
probability for Y to be thus distributed, obtained using the
Wilcoxon test, is 0.70 for ${\rm {W_r}^{min}=0.3\AA}$. Therefore
it is clear that redshift distribution of ${\rm Ly\;\alpha}$
clouds described by the equation (1) is consistent with the
observations if one considers lines which are more than 8 Mpc
away from the QSOs.

\centerline{\bf 4. PROXIMITY EFFECT}

Tytler (1987) calculated the intensity of the intergalactic background
(IGBR), ${\rm J_\nu = 10^{-21.7}\;ergs\;cm^{-2}\;s^{-1} \;Hz^{-1}\;
Sr^{-1}}$ at the Lyman limit based on QSO evolution studies, after
accounting for absorption by intervening Lyman limit systems (LLS).  He
concluded that photoionization by QSOs alone can not explain the
deficit of lines near the QSO unless the calculated background value
has been over estimated by a factor $\sim$ 60.  BDO constructed the
photoionization models of clouds near a QSO. Their model assumes the
shape of the ionizing background radiation and that of the QSO
continuum to be same.  For a given cloud near QSO, the H I column
density is given by

\begin{equation}
{\rm N_{H\;I}\; =\; N_o\;(1+\omega)^{-1},}
\end{equation}
where ${\rm N_o}$ is what the column density would have been if
there had been no nearby QSO and

\begin{equation}
{\rm \omega \;=\;{{F_\nu}^Q\over{4\pi J_\nu (z)}}}.
\end{equation}

${\rm J_\nu (z)}$ is the IGBR at the Lyman limit, at redshift z,
with which the clouds are in photoionization equilibrium and
\begin{equation}
{\rm {F_\nu}^Q\;=\;{L_\nu\over {4\pi {r_L}^2}}},
\end{equation}
is the local Lyman limit flux density due to the QSO, ${\rm
r_L}$ being the luminosity distance of the cloud from the QSO.
If the distribution of neutral hydrogen column density, ${\rm
{N_{H\;I}}}$  in individual clouds is a power law with index
$\beta$, i.e. if

\begin{equation}
{\rm {dN\over {dN_{H\;I}}}\propto {N_{H\;I}}^{-\beta}}
\end{equation}
then the number density of clouds above a column density
threshold, ${\rm N_o}$, is given by
\begin{equation}
{\rm N(N_{H\;I})\propto {N_o}^{1-\beta}}.
\end{equation}

Hence, for a sample limited by lower column density ${\rm
N_{H\;I}}$, the distribution of lines with redshift in the
spectra of a single QSO, including the proximity effect is
\begin{equation}
{\rm {dN\over dz}\propto
(1+z)^{\gamma}[(1+\omega(z))]^{1-\beta}}.
\end{equation}

BDO used ${\beta}$ = 1.7, obtained from high resolution
observations.  They applied this equation to equivalent width
limited samples assuming the following 3 idealizations. (1) the
column density distribution, with ${\beta = 1.7}$ holds for
clouds with column densities corresponding to equivalent widths
greater than ${\rm {W_r}^{min}}$.  (2) column densities and
velocity widths are uncorrelated (3) QSO proximity has no
significant effect on velocity width.  They obtained the IGBR to
be, ${\rm J_\nu = 10^{-21.0\pm 0.5}\;ergs\;cm^{-2}\;s^{-1}
\;Hz^{-1}\; Sr^{-1}}$ by fitting equation (7) to the observed
data. Lu et al (1991), applying the same model to the extended
low resolution data also got the same value of ${\rm J_\nu}$.
Recently Bechtold obtained ${\rm J_\nu =
10^{-20.5}\;ergs\;cm^{-2}\;s^{-1}
\;Hz^{-1}\; Sr^{-1}}$ with her intermediate resolution
data sample. Giallongo et al. (1993) based on column density
limited sample consisting of 3 QSOs got the value of ${\rm J_\nu
= 10^{-21.15}\;ergs\; cm^{-2}\;s^{-1}\;Hz^{-1}\; Sr^{-1}}$; they
used ${\beta = 1.53}$ and ${\gamma = 2.21}$. This value of ${\rm
J_\nu}$ is less than the values obtained based on equivalent width
limited samples,  validity of which depends on the 3 assumptions made
by BDO.  Also this value is only two times  larger than the
value obtained from QSO number density evolution models.
However results  for high resolution sample consisting of 8 QSO
spectra (paper 1) show that ${\rm J_\nu}$ has to be higher than
$10^{-21}$.

Here we apply the models of BDO to our sample considering only
the QSOs (marked by a * in Table 1.)  which do not show LLS within
8 Mpc  and do not show broad absorption lines. Any such line
may shield the QSO radiation and may complicate the analysis of
proximity effect. We used $\beta\; = \;1.7$, ${\rm q_o}=1/2$,
${\rm {W_r}^{min} =0.3\AA}$, and calculated the expected number
of lines for various relative velocity bins near QSOs for
various assumed values of ${\rm J_\nu}$.  We used ${\gamma}$
value obtained above. The value of ${\rm J_\nu}$ obtained by
minimizing $\chi^2$ is  ${\rm J_{\nu} = 10^{-20.22}\;ergs\;
cm^{-2}\;s^{-1}\;Hz^{-1}\; Sr^{-1}}$.  The probability that the
predicted distribution, with best fit value of ${\rm J_\nu}$,
represents the actual distribution is only 68${\%}$. For
Bechtold's value of ${\rm J_\nu}$ the probability falls to
61${\%}$. For the value of IGBR calculated by Madau
(1992) the probability is as low  as ${10^{-6}}$. The value of
of IGBR  obtained here is much higher than the one
expected from QSOs alone, and slightly higher than the previous
estimates from proximity effects.

Due to proximity effect, the expected number ${\rm n_e}$ of Ly
$\alpha$ lines near the QSOs, calculated from the general
redshift distribution is expected to be more than their observed
number ${\rm n_o}$. Therefore Y(= ${\rm n_e-n_o}$), is expected
to be distributed asymmetrically about zero. We have analysed
this and have looked for the presence of any correlation between
QSO properties with Y, for ${\rm {W_r}^{min}=0.3\AA}$. One
sample Wilcoxon test for lines within 8 Mpc and within ${\rm
r_{eq}}$  (as defined by Tytler (1987)), which is the distance
from the QSO at which the flux of radiation from QSO equals that
due to the IGBR, for ${\rm J_\nu =
10^{-21}\;ergs\;cm^{-2}\;s^{-1}
\;Hz^{-1}\; Sr^{-1}}$ shows the
probability  for Y to be normally distributed around zero to be
0.49 and 0.44 respectively.  We performed nonparametric Spearman
rank correlation test in order to find any possible correlation
between the properties of the QSO and Y. We do not find any
correlation between Y and the properties of QSO such  as, ${\rm
{z_{em}}^c}$, ${\rm f_\nu}$, ${\alpha_o}$ and RL.

It is interesting to note that there is no correlation between Y
and ${\rm f_\nu}$. This is against our expectation if the
proximity effect is induced by the excess ionization due to QSOs.
Note that Lu et al.(1991) also do not find any correlation between
luminosity, emission redshift or radio properties of the QSO and
the proximity effect. Bechtold (1994) did not find proximity
effect to depend on the radio properties or redshift but
qualitatively found a dependence on the luminosity of the QSOs.

Foltz et al. (1986) found an excess of C IV systems near the
QSOs over that predicted by the distribution of C IV systems far
from QSOs. The excess was shown to occur more frequently in the
radio loud objects.  These systems were attributed to absorption
due to the galaxies in the cluster of which QSO is also a
member. If the Ly ${\alpha}$ clouds are in some way associated
with galaxies then there will be excess absorption near the QSO
compared to that predicted by the general redshift distribution,
more so for the radio loud QSOs.  The lack of correlation
between radio loudness and the proximity effect can not,
however, be taken to indicate that there is no tendency of radio
loud QSOs not to show associated ${\rm Ly\;
\alpha}$ absorption as we are not considering QSOs with LLS within
8 Mpc of the emission redshift. Thus in a way we tend to choose
QSOs which are not having associated absorption and therefore
are not  residing in clusters.

\noindent{\bf 4.1  Effect of uncertainties in column density
distribution}

The column density distribution of Ly ${\alpha}$ lines in
individual cloud is not well determined due to the scarcity of
high resolution data.  Earlier studies by Carswell et al (1984)
obtained that the value of $\beta\sim 1.7$. Recent analysis of
Giallongo et al.  (1993) showed that ${\beta = 1.74}$ fits the
data with only 2{\%} probability and  for the lines with column
density between ${10^{13.2}}$ and ${10^{14.8}}$ the best fit
powerlaw is $\beta =1.53$ with 70{\%} probability.  They also
showed that the distribution is steeper for column density
greater than $10^{14.8}$.  In the case of 1331+170, Kulkarni et
al (1994) obtained $\beta = 1.5$ for 13.1$<{\rm log(N_{H\;I})}
<$13.9, and ${\beta = 3.1}$ for 13.9$<{\rm log(N_{H\;I})}
<$14.4. In order to consider the effect of variation in $\beta$
on the background intensity we varied the value of ${\beta}$ and
calculated $\chi^2$ probability for various values of ${\rm
J_\nu}$. The results are shown in fig 1.  Also given in the figure are
the expected values of the background intensity based on QSO
counts for various forms of QSO evolution and intergalactic
absorption (Madau 1992)

It is clear that the value of ${\rm J_\nu}$ decreases with
decrease in $\beta$, as has been earlier noted by Chernomordic
and Ozernoy (1993); also the probability that the observed
distribution is consistent with the predicted distribution
increases.  It is interesting to note that $\beta\;\sim 1.2-1.3
$ will not only make the value of ${\rm J_\nu}$ consistent with
the one expected from QSO counts, but also will make the
probability that the observed distribution is reproduced by the
predicted distribution very high.  Chernomordic and Ozernoy
(1993) showed, using the analytic equation for curve of growth,
that, in order to reproduce the observed equivalent width
distribution the ${\beta}$ value should be 1.4 rather than 1.7.
In their calculations they assumed the value of velocity
dispersion, b, to be 35 km s$^{-1}$, the mean value obtained
from the high resolution observations. Assuming the b value of
25 km s$^{-1}$ will reduce $\beta$ below 1.3.

\noindent{\bf 4.2 Effect of low resolution}

The sample used for this analysis is obtained using low
resolution spectra of QSOs. Most of the lines listed are
actually blends of few narrow lines. In the case of metal line
systems (Petitjean {\&} Bergeron, 1990) the number of components
is found to be correlated with the equivalent width of the
blended line. Jenkins (1986) showed that large number of
interstellar lines can be analysed collectively using the
standard, single component curve of growth. As long as the
equivalent widths of many line are combined and the distribution
function for the line characteristics is not markedly irregular,
one obtains nearly correct answer for the total column density
even when different lines have large variation in their central
optical depth and internal velocity dispersion. Also, the value
of velocity dispersion, b, obtained from standard curve of
growth of the blend will be higher than the one obtained by
profile fitting of high resolution data and need not reflect any
kinematic properties of the cloud, but is proportional to the
number of clouds( Srianand {\&} Khare, 1994). Thus one can write
\begin{equation}
{\rm N_{tot} = N_{tot}(1+\omega)^{-1},}
\end{equation}

${\rm N_{tot}}$ being the total column density of lines in the
blend. Thus the form of the equation used by BDO is still valid,
except that one has to use the distribution of total column
density including blends, instead of the distribution of column
density in individual components obtained from high resolution
observations. Blending of lines in low resolution data may occur
due to chance proximity of lines to each other but may also be
enhanced due to clustering of lines. Webb (1987) had found weak
clustering among Ly ${\alpha}$ lines. Recently we (Paper I) also
confirmed this from extended high resolution data. Barcons and
Webb (1990) showed that the discrepancy between the column
density distribution observed in high resolution data and that
obtained from the equivalent width distribution at low
resolution can not be accounted for by blending alone but can be
explained by invoking clustering. This effect further
invalidates the use of column density distribution obtained from
high resolution data in the proximity effect calculations using
low resolution data.

One can get the distribution of total column density, described by
${\beta_{\rm LR}}$, from the
observed equivalent width distribution through standard single
cloud curve of growth, with some assumed values of effective
velocity dispersion, ${\rm{b_{eff}}}$.  This is given by

\begin{equation}
{\rm f(N_{H\;I}) dN\;=\; f(W) {dW\over dN_{H\;I}}dN_{H\;I}}
\end{equation}

Assuming the standard forms for equivalent width and column
density distribution one can get $\beta_{\rm LR}$ from ${\rm
W_*}$ for a given value of ${\rm b_{eff}}$. We have determined
${\rm \beta_{LR}}$ from the observed equivalent width
distribution of our sample for different values of ${\rm
b_{eff}}$ restricting to the observed range of equivalent
widths.  $\beta_{\rm LR}$ values obtained for ${\rm W_*}=0.3\AA$
are given in table 3, with corresponding column density range
obtained and the significance of the fit.

It is clear from that table that the value of $\beta_{\rm LR}$
decreases with the decrease in velocity dispersion. Though our Ly
$\alpha$ sample contains strong lines, it does not contain Lyman
limit clouds, which means the ${\rm N_{tot}<10^{17}\;cm\;
s^{-2}}$.  Note that blends of several clouds in a small
velocity range $\sim 200\; {\rm km\;s^{-1}}$ can produce a Lyman
discontinuity if the total column density exceeds ${\rm \sim
2\times 10^{17}\;cm\; s^{-2}}$.  This condition ensures that the
effective velocity dispersion of the blend is higher than 40 km
s$^{-1}$.  The high resolution profile fitted results show that
the column density in individual clouds varies between $10^{13}
- 10^{16} {\rm cm ^{-2}}$. Blending will effectively
increase the column density cutoff values. Therefore the upper
limit will be $\geq 10^{16} {\rm cm^{-2}}$, and our ${\rm
b_{eff}}$ should be able to produce this range of ${\rm
N_{tot}}$. It can be seen from the table that in order to
satisfy the above conditions the ${\rm b_{eff}}$ values should
be between 40
and 50 ${\rm km\;s^{-1}}$. Which means that the $\beta_{\rm LR}$
is between 1.4 -1.6.

One can in principle obtain the value of effective velocity
dispersion using Ly $\alpha$ and Ly $\beta$ lines and using
doublet ratio method.  This value, however, will be biased
towards saturated lines, i.e. those which can produce
measurable amount of Ly $\beta$. Hunstead et al.(1988) and
Levshakov (1992) used Ly $\alpha$ and Ly $\beta$ lines to
calculate the effective column density of the blends. The
observed column density range in these calculations is ${\rm
10^{14.3}-10^{16.8}}$ with average velocity dispersion around 58
km s$^{-1}$, which is somewhat higher than the range obtained
above. However the average value of equivalent width of Ly
$\alpha$ lines used in their analysis is 0.78 $\AA$ which is
higher than the average equivalent width , 0.58
$\AA$, in our sample. The lower equivalent width blends most likely have
smaller ${\rm b_{eff}}$ unless they are formed due to blending
of very large number of weak lines. Thus the range in velocity
dispersion seems to be consistent with results of Hunstead et
al(1986) and Levshakov (1992).
Note that Press and Rybicki (1993) also found the value of
$\beta\;\sim$1.4 from equivalent width ratios of Lyman series lines. Their
value is independent of b. The equivalent width ratios were obtained by
considering the depressions in the continuum, without resolving
individual lines and are not biased.

Using $\beta$ values in the range obtained here instead of
$\beta$ =1.7, the value of ${\rm J_\nu}$ can at most be reduced
by a factor of 2.5 ( see fig 1.) but is still much higher than the
value predicted from QSO counts alone.

The total equivalent width of a metal line like Mg II or C IV is known
to be proportional to the number of components blended in the line
(York et al. 1986; Petitjean {\&} Bergeron 1990).  Srianand and Khare
(1994) showed that in order to explain the relationship between doublet
ratio and W, for Mg II doublets, with single cloud curve of growth, one
has to assume higher b values for lines with higher equivalent width.
In paper I we showed that the column density distribution of lyman
alpha lines in the blends is similar to the column density distribution
of the rest of the lines.  Thus strong lines are likely to have more
components and we expect ${\rm b_{eff}}$ to increase with W. In
addition there may also be an intrinsic tendency of b to increase with
line strength for Lyman alpha lines (Pettini etal, 1990) though this
has not not been confirmed by others (Carswell etal, 1991).If this kind
of correlation exists in the case of blended Ly $\alpha$ clouds it will
affect our calculation of ${\rm J_{\nu}}$.

We used Hunstead et al.(1986) and Levshakov (1992) data to
search for any such correlation. A clear correlation exists as
shown in figure  2 and is described by,
\begin{equation}
{\rm b=43.15\times W +31.38 \; km\;s^{-1}}
\end{equation}
Taking into account this correlation,we again calculated
${\beta_{\rm LR}}$ for ${\rm W_* = 0.3\AA}$, with ${\rm
b_{eff}}$ values given by the equation (10).  $\beta_{\rm LR}$
was found to be, 1.87, which gives a background value higher
than that estimated with ${\beta_{\rm LR}}$ = 1.7 by a factor of
1.5.  Note that the above relation between W and ${\rm
b_{eff}}$  is valid for high equivalent lines and may not
necessarily hold for the whole range of W.  However if valid,
the correlation between b$_{\rm eff}$ and W will widen the
difference between the values of ${\rm J_\nu}$ calculated from
proximity effect and from QSO counts.

\noindent{\bf 4.3 Effect of dust obscuration due to intervening
damped Ly $\alpha$ systems}

A series of studies conducted by Fall and his collaborators (
Fall, Pai {\&}  Mcmohan, 1989; Pai, Fall {\&} Bechtold, 1991)
indicate that the QSOs having intervening damped Ly $\alpha$
systems along the line of sight are redder than the rest of the
QSOs.  Recent studies by Pettini et al (1994), have confirmed
the existence of dust in the damped  Ly $\alpha$ systems. Though
the fraction of dust to gas ratio is not as high as that in the
case of the Milkyway, it does produce appreciable amount of
reddening in the QSO spectra.  The ${\rm f_\nu}$ values
calculated for these QSOs may thus be underestimates. IGBR
calculations from proximity effect, considering only
QSOs having damped Ly $\alpha$ absorbers are therefore expected
to yield smaller values for the background radiation field
compared to the values obtained using the whole sample. In order
to check this, we considered only QSOs which show intervening
damped Ly $\alpha$ systems and candidates (marked by d in Table
1) from our sample using the damped Ly $\alpha$ information
given in Lanzetta et al (1991). There are 16 such QSOs in our
sample. The value of ${\rm J_\nu}$ is obtained to be
$10^{-21}\;{\rm ergs\;cm^{-2}\;s^{-1}
\;Hz^{-1}\; Sr^{-1}}$.  This is smaller than the value obtained
for the whole sample by a factor of 3.  Our results thus  confirm
the reddening of QSOs due to intervening absorbers. This effect
causes  an error of upto a factor of 3 in the calculated ${\rm
J_\nu}$ values.

\centerline{\bf 5. PHYSICAL PROPERTIES OF Ly ${\alpha}$ CLOUDS NEAR
QSOs}

In this section we search for any possible changes in physical
properties of Ly ${\alpha}$ clouds like the equivalent width
distribution, column density distribution, distribution of velocity
dispersion and the distribution of expected and observed number of
lines near the QSOs.  Understanding how these properties change in the
vicinity of the QSOs will yield a better understanding of the proximity
effect.

\noindent{\bf 5.1 Equivalent width distribution}

We have calculated the equivalent width distribution of Ly
${\alpha}$ lines within 8 Mpc of the QSO for various values of
${\rm {W_r}^{min}}$.  The values of ${\rm W_*}$ are very similar
to the values of ${\rm W_*}$ for lines away from the QSOs, given
in table 2.  KS test confirms that the two distributions are
drawn from the same parent population.  This is consistent with
the results of earlier analysis (Sargent et al. 1980; Lu et al,
1991; Bechtold, 1994).  However it is possible that the
proximity effect does not extend to 8 Mpc and any change in the
equivalent width distribution may show up if we restrict to
regions within '${\rm r_{eq}}$'. We have calculated the values
of ${\rm W_*}$ for various values of ${\rm {w_r}^{min}}$ and of
IGBR for lines within ${\rm r_{eq}}$ from the
QSOs.  The results are shown in figure 3. It is clear from the
figure that the ${\rm W_*}$ increases for higher values of
background intensity. This effect is more pronounced for the
higher equivalent width lines.  This suggests that the ratio of
number of strong lines to that of weak lines increases near the
QSO.

Assuming ${\rm J_\nu\;=\;10^{-21}\;ergs\;cm^{-2}\;s^{-1}
\;Hz^{-1}\; Sr^{-1}}$ we calculated ${\omega}$,
the ratio of ionizing flux from the QSO and the background
radiation flux, for each line within ${\rm r_{eq}}$.
Nonparametric Spearman rank correlation test was performed in
order to find any correlation or anticorrelation between W and
${\omega}$. No significant correlation was found for ${\rm
{W_r}^{min}=0.16\AA}$ and 0.3$\AA$. However there is 2.6$\sigma$
correlation between the two quantities if we consider lines with
equivalent width greater than 0.6$\AA$. The null hypothesis that
the two quantities are uncorrelated can be rejected at more than
99.6{\%} confidence level. This again shows that  strong lines
tend to occur more frequently compared to weak lines near the
QSOs.  We thus conclude that the equivalent width distribution
is affected by the excess photoionization due to QSO. This may
be due to increase in velocity dispersion in individual
components or may be due to the effect of enhanced blending
due to excess clustering of Ly $\alpha$ around the QSOs

\noindent{\bf 5.2 Column density distribution}

The change in the ${\rm W_*}$ near the QSOs suggests that the column
density distribution may also be different in regions close to the
QSOs.  We therefore calculated the column density distribution  of Ly
${\alpha}$ lines for various values of ${\rm {N_{H\;I}}^{min}}$,
considering lines within ${\rm r_{eq}}$ only for the high resolution
data compiled in paper1, adding lines from Fan and Tytler (1994) for
QSO 1946+7658.  The results for various assumed values of ${\rm J_\nu}$
are shown in figure 4. It can be seen from the figure that the error
bars are too large to reveal any change in ${\beta}$.  If any, there
may be a slight decrease in $\beta$ with increasing ${\rm J_\nu}$.
Spearman rank correlation test between ${\rm N_{H\;I}}$ and ${\omega}$
for ${\rm J_\nu= 10^{-21}\;ergs\;cm^{-2}\;s^{-1} \;Hz^{-1}\; Sr^{-1}}$
for various ${\rm {N_{H\;I}}^{min}}$ values does not show any
significant correlation or anticorrelation between the two quantities.
Therefore though figure 4 seems to indicate a slight decrease in
${\beta}$ the rank correlation test does not show any statistically
significant change in column density with ${\omega}$ which is
consistent with the assumption of BDO that the column density
distribution is unaltered by the presence of QSO.

\noindent{\bf 5.3 Distribution of velocity dispersion parameter}

The  equivalent width and column density distributions near the QSOs
suggest that the velocity dispersion parameter in clouds near QSOs may
be higher. If the reported correlation between column density and b is
true then also one would expect an increase in velocity dispersion due
to proximity effect.  Similar result is also expected on the basis of
ionization models. In figure 6 we have plotted the distribution of b
for clouds within 8 Mpc for ${\rm J_\nu=10^{-21}\;ergs\;cm^{-2}\;s^{-1}
\;Hz^{-1}\; Sr^{-1}}$ for $\omega>1 \;{\rm and}\; \omega <1$
separately. It is clear that low values of b, (i.e. b $<20\;{\rm
km\;s^{-1}})$ observed in $\sim 25{\%}$ of the cases for $\omega<1$ are
not seen in the case of $\omega>1$. The clouds on an average have
higher 'b' values when $\omega>1$.  Spearman rank correlation test
between $\omega$ and b for ${\rm J_\nu =
10^{-21}\;ergs\;cm^{-2}\;s^{-1} \;Hz^{-1}\; Sr^{-1}}$ shows 2.1$\sigma$
correlation between $\omega$ and b. The null hypothesis that the two
quantities are uncorrelated can be rejected at 96{\%} confidence level.
An increase in b will increase the ${\rm b_{eff}}$ and therefore will
give higher value for ${\rm J_\nu}$.

We thus conclude that the column density distribution seems to be
similar for lines near as well as away from the QSO, while the velocity
dispersion parameters and as a result  the equivalent width distribution
is different for the two classes of lines.

\centerline{\bf 6. PROPERTITS OF NVLs}

More than 50{\%} of the QSO sight lines in our sample show
narrow absorption features in the red wing of the Ly $\alpha$
emission which can not be attributed to any strong UV absorption
belonging to previously known absorption systems listed by
Junkkarinen et al.(1991). These are believed to be Ly ${\alpha}$
lines redshifted with respect to the QSO. Note that metal line
systems with ${\rm z_{abs}>z_{em}}$ have also been seen and are
believed to be associated with the QSOs. They probably occur
more frequently in radio loud or intrinsically faint QSOs (Foltz et al
1986; Moller {\&} Jakobson, 1987).

In order to determine if the occurrence of the NVLs depends on
any of the QSO properties, we first looked for the correlation
between the presence of NVLs and the presence of associated
metal lines, i.e. metal line systems within 8 Mpc of QSOs. Out
of 15 QSOs not having NVLs only two show associated metal line
absorption, whereas for 28 QSOs having at least one NVL 15 show
metal line absorption within 8 Mpc. The occurrence of metal line
systems near QSO and the occurrence NVLs thus seem to be
correlated at about 7.5$\sigma$ level. However we do not find any
correlation between the
number of associated metal line systems and the number of NVLs.

Next, we checked the possible correlation between the QSO flux
at the Lyman limit, ${\rm f_{\nu}}$ and the occurrence of NVLs.
We did not find any correlation between these two quantities.
There is no difference in the distribution of ${\rm f_{\nu}}$
for QSOs which show NVLs and that for the rest of the QSOs.

The occurrence of  NVLs also seems to  be uncorrelated with the
radio properties of QSOs.  Out of 18 QSOs for which radio
information is available, 2 radio quiet QSOs show more than 3
NVLs, 9 radio loud QSOs show more than one NVL and the rest 7
radio loud QSOs do not show any NVLs. We also do not find any
correlation between optical spectral index , ${\alpha}$ and the
occurrence of NVLs.  The distribution of optical spectral index
for QSOs with and without NVLs are identical.

Next, we divided our sample into  various redshift bins and
calculated the ratio of the number of QSOs showing NVLs and the
number of QSOs which do not show such lines, in each of these
bins. The results are shown in figure 6. There is a clear
increase in the ratio with redshift, implying that  high
redshift QSOs show NVLs more frequently than their low redshift
counterparts.  Also, there seems to be a weak trend of
increasing number of NVLs in the spectra of an individual QSO
with redshift of the QSO. The fraction
of QSOs having a given number of NVLs is plotted in figure 7 for
two different redshift bins.

In the framework of cosmological redshifts as distance
indicators, we consider two possible origins for these lines,
(1) The emission redshift of the QSOs used here may be wrong and may
actually be higher than the values used here, (2) The redshifts
are not the correct indicators of distance and peculiar
velocities of the Ly ${\alpha}$ clouds or the QSOs can be the
cause for NVLs Below we consider these possibilities and try to
estimate the IGBR in each case.

\noindent{\bf 6.1 Effect of emission line velocity shifts}

In this section we consider the possibility that the observed Ly
${\alpha}$ lines with ${\rm z_{abs}>z_{em}^c}$ are due to the
underestimation of the QSO redshifts, and try to understand the
results discussed in the previous section and the proximity
effect in this scenario. Such a possibility has been considered
by Espey (1993) who showed that a bias in the measurement of
QSO redshifts can account for at least a portion of the
discrepancy between the predicted values of the ionization
background and the values resulting from the known QSOs. Note
that the emission line redshifts used by us are obtained by
using all observed emission lines and the analytic equations
given by Tytler and Fan (1992). However these formulae assume a
simple extrapolation of known results based on intermediate
redshift QSOs, where both low as well as the high ionization
lines are observable, to high redshifts, where most often the
redshifts are based solely on the high ionization lines.
Corbin (1990) showed that there is a clear correlation between
v(Mg II - C IV), the velocity difference between the Mg II and C
IV emission lines, and absolute magnitude of the QSO.  This was
also confirmed by Tytler and Fan (1992). If Corbin relationship
holds good for
all redshifts then, due to the known luminosity evolution of
QSOs (Boyle et al. 1988), the velocity shifts will be higher for
higher redshift QSOs, and corrections applied by us may still be
less than the actual offset values.

Following Espey (1993) we shifted all QSO redshifts by , ${\rm
v_s}$, and calculated the expected number for ${\rm
Ly_{\alpha}}$ lines in various relative velocity bins for ${\rm
{W_r}^{min} = 0.3\AA}$.  Espey (1993) considered only lines with
${\rm z_{abs}}$ less than the uncorrected emission redshift,
${\rm {z_{em}}}$ for the completeness of the sample.  However as
we have information beyond ${\rm {z_{em}}}$ for several QSOs, we
consider all lines with ${\rm z_{abs}}$ less than the corrected
emission redshift, ${\rm {z^s}_{em}}$, after applying the shift,
${\rm v_s}$. Inclusion of these lines will increase the size of
the sample and should therefore give better (with smaller errors)
values of the IGBR in this scenario.

The best fit values of ${\rm J_\nu}$ obtained by minimizing
$\chi^2$, for various values of ${\rm v_s}$ and the probability for
various values of ${\rm v_s}$ to describe the observations are
given in Table 4.  Increasing ${\rm v_s}$ decreases the value of
IGBR however, the probability that the model reproduces the
observations decreases. This is because large shifts predict
more than the observed number of lines with ${\rm
{z^s}_{em}<z_{abs}<{z_{em}}}$.  It thus appears that
${\rm v_s}$ has to be less than 1500 ${\rm km\; s^{-1}}$ in
order to reproduce the observed distribution of Ly $\alpha$
clouds near the QSOs.  There are, however, Ly ${\alpha}$ lines
redshifted with respect to QSOs by more than 2500 km s$^{-1}$,
which implies that at least in few QSOs the correction to the
emission redshift may be much higher than the upper limit
obtained from our analysis.  Systemic corrections $\sim 5000 \;
{\rm km\;s^{-1}}$ have been obtained by Bechtold (1994) in few
cases.  In cases like 1331+170, however lines are seen upto a
relative velocity of ${\rm \sim -10,000 \; km\;s^{-1}}$
(Kulkarni et al, 1994) and may require alternative explanation.
The observed distribution of lines near the QSO and the distribution
predicted for ${\rm v_s\;=\;1000\;km\;s^{-1}}$ is shown in figure 8
for ${\rm J_\nu\;=\;10^{-21.0}\;ergs\;cm^{-2}\;s^{-1}
\;Hz^{-1}\; Sr^{-1}}$

According to Corbin correlations we should expect large
corrections to redshifts for more luminous objects and hence
more number of NVLs, in these QSOs. However more luminous QSOs
will also have stronger proximity effect, thereby reducing the
number of such lines. The absence of any correlation or
anticorrelation between the number of NVLs and ${\rm f_\nu}$ or
${\alpha}$ can thus be understood as the combined effect of
Corbin correlations and proximity effect.  Another trend found
by us is that more number of NVLs tend to occur at high
redshifts.  This can be understood as an increase in the shift,
${\rm v_s}$, with redshift for a constant background or an
increase in IGBR with increasing redshift.

\noindent{\bf 6.2 Effect of peculiar velocities }

Another explanation for NVLs may be the presence of large
peculiar velocities of the intervening Ly $\alpha$ clouds on top
of the general Hubble flow. These peculiar velocities may either
be due to the gravitation induced clustering of clouds
associated with collapsed density peaks as in the models of
Rees (1986) or due to the origin of clouds in the fragmenting
shells as in the models proposed by Ostriker and Ikeuchi (1983).
If Ly $\alpha$ clouds are formed in the extended galactic halos,
in extended disks or in dwarf galaxies, one would expect them
to have some peculiar velocity on top of the Hubble flow.  It is
also possible that QSOs themselves have peculiar velocities. In
what follows we assume that all the Ly $\alpha$ clouds have
peculiar velocities. We neglect the peculiar motions of QSOs
keeping in mind that neglecting the peculiar velocities of Ly
$\alpha$ clouds and considering peculiar motion of QSOs alone is
identical to what is considered here as far as the proximity
effect is concerned. The distribution of the line of sight
component of the peculiar velocities of Ly $\alpha$ clouds is
assumed to be gaussian given by,
\begin{equation}
{\rm f(v) = {1\over{\sqrt{\pi}v_d}} \exp
\bigg[-\bigg({v\over{v_d}}\bigg)^2\bigg]},
\end{equation}
where ${\rm v_d}$ is the dispersion in peculiar velocity.
Taking into account the I-model discussed by BDO, one can write
the number of absorption lines expected at any redshift z along
the line of sight to a QSO to be,
\begin{equation}
{\rm N(z)={N_o\over {\sqrt{\pi} v_d}}\int\limits_0^{z_{em}^c}
(1+z_1)^{\gamma +1} (1+\omega)^{1-\beta} \exp \bigg[-{v(z,z_1)^2\over
v_d^2}\bigg] dz_1},
\end{equation}
where
\begin{equation}
{\rm {v(z,z_1)\over c} =
{{(1+z_1)^2-(1+z)^2}\over{(1+z_1)^2+(1+z)^2}}},
\end{equation}
c being the velocity of light in km s$^{-1}$. Taking the value
of ${\beta}$ = 1.7 and the value of $\gamma$ obtained above we
calculated the expected number of lines in bins with different
relative velocities with respect to QSOs for various values of
${\rm v_d}$ and ${\rm J_\nu}$.

Considering only bins between -2500 and 6000 km s$^{-1}$, the
$\chi^2$ probability that the predicted number of lines is
consistent with the observed values are given in Table 5.
The expected distribution for ${\rm v_d\;=\;2000\;km\;s^{-1}}$ and
${\rm J_\nu\;=\;10^{-20.5} ergs\;cm^{-2}\;s^{-1}
\;Hz^{-1}\; Sr^{-1}}$ is shown in figure 8. It is
clear from the table that ${\rm J_\nu}$ has to be higher than
${\rm 10^{-20.5}\;ergs\;cm^{-2}\;s^{-1}
\;Hz^{-1}\; Sr^{-1}}$ with ${\rm v_d}$ around 2000 km s$^{-1}$
in order to fit the observed distribution. This value of peculiar
velocity is very large. Dressler et al.'s (1987) analysis
indicates that the peculiar velocities of galaxies are about 600
km s$^{-1}$ relative to the microwave background. However the
peculiar velocities of clusters of galaxies are around 2000 km
s$^{-1}$ (Bahcall, 1988).

If the high values of peculiar velocities are real then one would
expect to get excess correlation upto the scale of few thousand km
s$^{-1}$ in the pair velocity correlation function of Ly $\alpha$
lines.  However, studies conducted by various groups do not show any
such excess beyond 300 km s$^{-1}$. It is possible that the NVLs may be
Lyman alpha clouds associated with the galaxies in the cluster of
galaxies containing the QSO. Presence of such a population of Lyman
alpha clouds, distinct from the intergalactic Lyman alpha clouds, is
indicated by the HST observations (Lanzetta etal 1995). This is also
supported by the correlation of the NVLs and associated absorption
systems, noted above. The peculiar velocities, in this case may be
confined to this population of Lyman alpha clouds and may not,
therefore, show up in the two point correlation of the entire
population. However the peculiar velocities of galaxies in cluster are
known to be only about 600 km s$^{-1}$ as noted above and unless the
peculiar velocities of the galaxies in the QSO containing cluster are
larger than those in other clusters, it is unlikely that the NVLs are
caused by the peculiar velocities alone. Also most of the associated
metal line systems which are formed by galaxies in such clusters have
metallicity close to or larger than the of solar value. If Ly
$\alpha$ clouds are members of the QSO cluster then they are
expected to be large in order to have detectable neutral hydrogen
and thus they would have accompanying high ionization metal
lines. Such lines have not been observed. It thus seems unlikely
that NVLs are produced by such absorbers. If these clouds are
highly ionized massive clouds then one should frequently see
x-ray absorption in such high z QSOs. Thus the x-ray
observations of high z QSOs can be utilized to put bound on the
fraction of lines that can be possibly present in the cluster
containing the QSO.

One can consider the possibility that the QSOs themselves have
peculiar velocities of $\sim 2000\; {\rm km\;s^{-1}}$ on top of the
Hubble flow. The values of ${\rm J_\nu}$ and ${\rm v_d}$
obtained above will be valid for this case also. In order to
have such a high velocity, QSO host clusters should have a mass
of about ${\rm\sim 10^{16}M_\odot}$ and more than 50 {\%} of the
QSOs should form in such massive cluster environments.  The
formation of such massive structures at redshifts as early as z
= 4 is difficult in the currently favored models of structure
formation.  We can thus rule out the possibility that NVLs are
produced due to peculiar velocities of Ly $\alpha$ clouds or
those of QSOs. Also if this scenario was correct it will widen
the discrepancy between the calculated background and that
expected from the QSO counts alone.

It is clear from the above analysis that there is no unique
mechanism by which one can explain the presence of NVLs. However
a systemic offset of 1000-1500 km s$^{-1}$ with a peculiar
velocity dispersion of 500 km s$^{-1}$ may explain the NVLs. The
background intensity calculated in such a model will be similar
to the values obtained by considering proximity effect for lines
with ${\rm z_{abs}<z_e}$ alone and the discrepancy between the
background intensity calculated by using proximity effect and
from QSO counts will persist.

\centerline{\bf 7. CONCLUSIONS}

A large sample of 69 QSOs,all observed with intermediate
resolution between 60 and 100 km s$^{-1}$ has been complied.
The sample contains 1671 lines with equivalent width ${>0.3\AA}$.
It is analysed for evolutionary properties and proximity effect.
The following conclusions are drawn.

(1) Ly $\alpha$ lines more than 8 Mpc away from QSOs  give
${\rm W_*\; =\;0.285\pm0.008}$ and
${\gamma\;=\;1.725\pm0.228}$.

(2) The analysis of proximity effect among Ly $\alpha$ lines
close to the QSO, gives ${\rm
J_\nu\;=\;10^{-20.22}\;ergs\;cm\;^{-2}\;s^{-1}
Hz^{-1}\;Sr^{-1}}$

(3) Value of J$_\nu$ obtained from proximity effect
 is shown to be sensitive to the assumed
column density distribution and decreases by a factor of 3 if
the powerlaw index of the distribution is changed from 1.7 to
1.4.

(4) It is proposed that the column density distribution,
obtained from the observed equivalent width distribution
assuming an effective velocity dispersion parameter should be
used in the I-model calculations of the proximity effect in an
intermediate resolution sample. This is shown to reduce the
necessary background flux by a factor of 2.5.

(5) A correlation between the effective velocity dispersion
parameter and equivalent width is shown to be likely to be
present. Such a correlation, if present, tends to increase the
background radiation flux.

(6) Considerations of proximity effect in the spectra of 16
QSOs, from our sample exhibiting damped Ly $\alpha$ lines, gives
a background intensity 3 times smaller than the value
obtained from the whole sample, confirming the presence of dust in the
damped Ly ${\alpha}$ systems.

(7) The equivalent width distribution of Ly $\alpha$ lines close
to QSOs is shown to be flatter than the distribution away
from QSOs.

(8) The average velocity dispersion parameter for lines near the
QSOs is shown to be higher by a factor of 1.2  than that for
lines away from the QSOs.

(9) The occurrence of Ly $\alpha$ lines with absorption redshift
larger than the emission redshift of the QSO is shown to be
uncorrelated with QSO luminosity, optical spectral index or
radio loudness. It  is shown to be correlated with the redshift
of the QSO as well as with the presence of associated metal
systems.

(10) The number of absorption lines with redshift higher than the
emission redshift in a QSO spectra is shown to increase with redshift.

(11) An underestimate of systemic redshift of QSOs by about 1500 km
s$^{-1}$ can explain the distribution of lines near QSOs including
NVLs. However the significance of the fit is shown to decrease as one
tries to fit the high velocity NVLs.  Though this scenario reduces the
IGBR by a factor of 2 to 3 it fails to explain the higher velocity
NVLs.

(12) Very high peculiar velocities $\sim2000\;{\rm km\;s^{-1}}$
of the Ly $\alpha$ clouds or the QSOs are needed to account for
the occurrence of NVLs assuming no underestimate of systemic
redshift. The
required values of IGBR is higher than that obtained without peculiar
velocities.

(13) Both the above effects namely higher systemic redshifts of
the QSOs and peculiar velocities of Ly $\alpha$ clouds (or QSOs)
together may explain the occurrence of NVLs.

 In this work we have highlighted some of the issues that one has to
consider while calculating the IGBR based on proximity effect.
Understanding the origin of NVLs is very important  as it plays
a vital role in the calculations of IGBR. It can also  provide
some important clues towards our understanding of the QSO properties
and the structures in the early universe. Only a large, column density
limited, samples obtained by profile fitting of high resolution data
can give a better estimation for the IGBR as
the curve of growth effects seem to produce an uncertainty of about
a factor of up to 5 in the calculated values of IGBR.

 We wish to thank Dr. V. Sahni, Dr. K. Subramaniam and Dr. B. S.
Sathyaprakash for useful discussions regarding peculiar
velocities. This work was supported by a grant No. SP/S2/021/90
by the Department of Science and Technology, Govt. of India.

\newpage

\centerline{\bf Figure Captions}

Fig 1. $\chi^2$ probability for various values of $\beta$ and ${\rm J_\nu
}$. Dotted line indicates the value of IGBR obtained by Madau (1992)
based on QSO number density.

Fig 2. Velocity dispersion parameter for individual blends vs. equivalent
width with the best fit line.

Fig 3. Equivalent width distribution parameter ${\rm W_*}$ for lines
within ${\rm r_{eq}}$ vs. ${\rm J_\nu}$.

Fig 4. Column density distribution parameter ${\beta}$ for lines
within ${\rm r_{eq}}$ vs. ${\rm J_\nu}$.

Fig 5. Distribution of velocity dispersion for $\omega<1$ and $\omega>1$.

Fig 6. Ratio of number of QSOs with and without NVLs as a function of
redshift.

Fig 7. Fraction of QSOs showing NVLs as a function of number of NVLs.

Fig 8. Histogram showing observed number of Ly $\alpha$ lines  as
a function of relative velocity with respect to QSO. Solid line shows
the observed distribution. Dotted line shows the expected number for
${\rm v_s\;=\;1000\;km\;s^{-2}}$ and ${\rm J_\nu\;=\;10^{-21}\;ergs\;
s^{-1}\;cm^{-2}\;st^{-1}\;Hz^{-1}}$. Dashed line shows the expected
number for ${\rm v_d\;=\;2000\;km\;s^{-2}}$ and ${\rm J_\nu\;=\;10^
{-20.5}\;ergs\; s^{-1}\;cm^{-2}\;st^{-1}\;Hz^{-1}}$.

\newpage
\widetext
\begin{table}
\caption {Data sample}
\begin{tabular}{cccccccccc}
\multicolumn{1}{c}{QSO}&\multicolumn{1}{c}{${\rm z_{em}}$}&
\multicolumn{1}{c}{${\rm {z_{em}}^c}$}&\multicolumn{1}{c}{${\rm z_{min}}$}&
\multicolumn{1}{c}{${\rm z_{max}}$}&\multicolumn{1}{c}{E}&\multicolumn{1}{c}
{${\rm f_\nu^+}$}&\multicolumn{1}{c}{$\alpha$}&\multicolumn{1}{c}{RL}&
\multicolumn{1}{c}{ref.}\\
\tableline
 0000-263*d&4.111& 4.114& 3.303& 5.580& 0.15& 586& 1.090& quiet&9\\
 0001+087*&3.243& 3.241& 3.035& 3.421& 0.14& 270& 0.290& .....&1\\
 0002+051*&1.899& 1.900& 1.764& 2.628& 0.16& 663& 1.010& 2.00 &2\\
 0002-422*&2.763& 2.767& 2.175& 3.418& 0.30& 301& .....& 0.87 &3,4\\
 0014+813*&3.384& 3.387& 2.743& 3.483& 0.15& 909& 1.600& 2.55 &1\\
 0029+073&3.294& 3.262& 2.869& 3.384& 0.14& 377& 1.070& .... &1\\
 0042-264*&3.298& 3.300& 3.192& 5.950& 0.15& 337& .....& .... &9\\
 0055-269*d&3.653& 3.663& 3.347& 5.909& 0.15& 177& 0.470& .... &9\\
 0058+019*d&1.960& 1.961& 1.789& 1.969& 0.13& 405& 0.800& 0.18 &5\\
 0114+089*&3.205& 3.163& 2.685& 3.434& 0.08& 317& 0.750& .... &1\\
 0142-100*&2.719& 2.728& 2.142& 2.922& 0.13& 658& .....& 0.60 &5\\
 0149+336*d&2.430& 2.432& 2.174& 3.195& 0.15& 191& 0.900& .... &12\\
 0201+365&2.912& .....& 2.298& 2.611& 0.15& ...& .....&. ....&11\\
 0237-233*&2.222& 2.225& 1.756& 3.121& 0.13& 844& 0.170& 2.51 &5\\
 0256-000*&3.374& 3.377& 2.685& 3.434& 0.18& 408& 0.580& .... &1\\
 0301-005&3.223& 3.226& 2.759& 3.212& 0.12& 280& 0.890& .... &1\\
 0302-003*&3.286& 3.290& 2.685& 3.434& 0.17& 459& 0.630& .... &1,8\\
 0334+204*d&3.126& 3.312& 3.043& 3.212& 0.06& 227& 0.650& .... &1\\
 0347-383*d&3.230& 3.233& 3.146& 5.909& 0.15& 248& 0.750& 0.90 &9\\
 0421+019*&2.051& 2.055& 1.772& 2.826& 0.16& 131& 0.850& 2.93 &2\\
 0424-131*&2.166& 2.168& 1.906& 2.581& 0.12 & 59& 1.500& 2.72 &5\\
 0428-136&3.244& .....& 3.162& 5.909& 0.20& ...& .....& .... &9\\
 0453-423*&2.656& 2.657& 2.209& 3.402& 0.32& 323& .....& 0.81 &3,4\\
 0636+680*&3.174& 3.178& 3.043& 3.212& 0.13& 1382& 1.200& 3.52&1\\
 0731+653*&3.033& 3.038& 2.677& 3.359& 0.30& 160& 0.660& 2.58 &1\\
 0831+128*&2.739& 2.737& 2.156& 2.825& 0.29& 161& 0.660& .... &1\\
 0837-109*&3.326& 3.329& 2.989& 3.442& 0.10& 178& .....& .... &5\\
 0848+163*&1.925& 1.926& 1.764& 2.251& 0.07& 108& 0.460& 1.20 &5\\
 0905+151*&3.173& 3.175& 3.043& 3.212& 0.07&  72& .....& .... &1\\
 0913+072*d&2.784& 2.786& 2.193& 2.825& 0.19& 495& .....& .... &1\\
 0938+119*d&3.192& 3.194& 3.043& 3.212& 0.12&  46& 0.320& 3.06 &1\\
 0953+549&2.584& 2.586& 2.018& 2.574& 0.15& 350& .....& .... &10\\
 0956-122*&3.306& 3.306& 2.964& 3.330& 0.15& 270& 0.490& .... &8\\
 1017+280*d&1.928& 1.929& 1.632& 2.661& 0.10& 1344& 1.040& ....&5\\
 1017-106&3.158& 3.158& 2.855& 3.222& ....& ...& .....& .... &8\\
 1033+137*&3.092& 3.094& 3.043& 3.212& 0.05& 253& .....& loud &1\\
 1115+080*&1.725& 1.728& 1.682& 2.489& 0.16& 648& 1.040& 0.60 &2\\
 1151+068*d&2.762& 2.764& 2.173& 3.216& 0.30& 130& 0.900& .... &7\\
 1159+124*&3.502& 3.504& 2.795& 3.607& 0.20& 881& 0.220& .... &5\\
 1206+119&3.108& 3.110& 2.784& 3.269& 0.09& 279& .....& .... &1\\
 1208+101*&3.822& 3.825& 3.400& 4.018& 0.11& 333& 1.260& .... &1\\
 1215+330*d&2.606& 2.617& 2.043& 2.784& 0.16& 154& 0.320& 2.32 &1\\
 1225+317*&2.200& 2.210& 1.698& 2.979& 0.16& 1169& 0.620& 2.20 &4\\
 1244+349*d&2.500& 2.502& 2.180& 2.714& 0.15&  69& 0.640 &.... &12\\
 1247+267*d&2.039& 2.042& 1.680& 2.455& 0.08& 573& 0.620& 0.60 &5\\
 1334+005&2.842& .....& 2.242& 2.783& 0.07& 560& .....& .... &1\\
 1400+114*&3.177& 3.179& 3.043& 3.212& 0.13& 115& .....& .... &1\\
 1402+044&3.206& 3.211& 3.032& 3.384& 0.30&  45& 0.740& 3.40 &1\\
 1409+093&2.912& .....& 2.236& 2.838& 0.30& ...& ....& ....  &11\\
 1410+096*&3.313& 3.316& 3.032& 3.384& 0.09& 134& .....& .... &1\\
 1442+101*&3.554& 3.535& 2.872& 3.582& 0.09& 278& 0.380& 3.38 &1\\
 1451+123&3.251& 3.257& 3.032& 3.384& 0.20& 157& .....& .... &1\\
 1512+132*&3.120& 3.122& 2.872& 3.228& 0.25&  64& .....& .... &1\\
 1548+092*d&2.748& 2.749& 2.455& 2.907& 0.29& 336& .....& .... &1\\
 1601+182*&3.227& 3.229& 3.032& 3.384& 0.22&  34& .....& .... &1\\
 1602+178*&2.989& 2.991& 2.749& 3.105& 0.40&  32& 1.220& .... &1\\
 1614+051*&3.216& 3.217& 3.032& 3.384& 0.25&  22& .....& 3.98 &1\\
 1623+269*&2.526& 2.527& 1.973& 3.623& 0.15& 413& 0.310& .... &5\\
 1738+350*&3.239& 3.240& 3.032& 3.384& 0.32&  25& 0.660& 3.09 &1\\
 2000-330*&3.777& 3.783& 3.113& 5.909& 0.14& 328& 0.750& loud &9\\
 2126-158*d&3.280& 3.267& 2.608& 4.179& 0.32& 577& 1.430& 3.22 &4\\
 2204-408*&3.169& 3.171& 3.129& 5.909& 0.15& 363& .....& .... &6\\
 2233+131*d&3.295& 3.298& 3.032& 3.384& 0.14& 235& 0.820& .... &1\\
 2233+136*&3.209& 3.216& 2.869& 3.384& 0.09& 199& 0.540& .... &1\\
 2239-386&3.511& .....& 3.442& 5.909& 0.30& ...&....&. ....  &9\\
 2311-036*&3.048& 3.048& 2.750& 3.121& 0.37 & 92& 0.680& .... &1\\
 2359-022&2.810& .....& 2.473& 2.578& 0.15& ...& .....& .... &12\\
\end{tabular}
\tablenote{ref: (1) Bechtold  (1994), (2) Young, Sargent {\&} Boksenberg
(1982), (3) Sargent et al.(1979), (4) Sargent et al. (1980), (5) Sargent,
Boksenberg {\&} Steidel (1988), (6) Willinger et al. (1989), (7) Turnshek
et al. (1989), (8) Steidel (1990b), (9) Steidel (1990a), (10) Levshakov
(1992),
(11) Lu et al. (1993), (12) Wolfe et al. (1993).\\
$*$ QSOs used for the proximity effect analysis\\
$d$ QSOs which have intervening damped Ly $\alpha$ absorbers\\
$+$ ${\rm f_\nu}$ is in units of microjanskies}
\end{table}

\newpage
\widetext
\begin{table}
\caption { Results of maximum likelihood analysis}
\begin{tabular}{cccccc}
\multicolumn{1}{c}{${\rm {W_r}^{min}}$}&\multicolumn{1}{c}{$\gamma$}&
\multicolumn{1}{c}{${\rm P_{KS}}$}&\multicolumn{1}{c}{${\rm W_*}$}&
\multicolumn{1}{c}{$\chi^2$}&\multicolumn{1}{c}{$<{\rm Q_K -0.5}>$}\\
\tableline
0.16&1.0624$\pm$0.1944&1e-22&0.2858$\pm$0.0079&1.66&0.0043\\
0.20&1.4161$\pm$0.2015&1e-11&0.2824$\pm$0.0076&1.29&-0.0005\\
0.25&1.7553$\pm$0.2270&0.95&0.2780$\pm$0.0080&1.34&0.0043\\
0.30&1.7245$\pm$0.2279&0.57&0.2850$\pm$0.0080&0.92&-0.0005\\
0.45&2.2480$\pm$0.3016&0.66&0.2840$\pm$0.0100&1.47&0.0102\\
0.60&2.1730$\pm$0.3880&0.65&0.2780$\pm$0.0130&1.48&0.0132\\
\end{tabular}
\end{table}
\newpage
\mediumtext
\begin{table}
\caption{Effective column density distribution}
\begin{tabular}{cccc}
\multicolumn{1}{c}{${\rm b_{eff}}$}&\multicolumn{1}{c}{range of N(H I)}&
\multicolumn{1}{c}{$\beta_{\rm LR}$}&\multicolumn{1}{c}{significance}\\
\tableline
30&14.2--18.0&1.1628&0.99\\
40&14.0--17.3&1.3769&0.99 \\
50&14.0--16.2&1.6035&1.00\\
60&13.9--15.5&1.8343&1.00\\
70&13.9--14.8&2.0803&1.00\\
\end{tabular}
\end{table}
\newpage
\mediumtext
\begin{table}
\caption{Results of proximity effect calculations taking into account
possible shift in
systemic redshifts}
\begin{tabular}{ccc}
\multicolumn {1}{c}{${\rm v_s}$}&\multicolumn {1}{c}{${\rm log{J_\nu^*}}$}&
\multicolumn{1}{c}{${\rm \chi^2\;(prob)}$}\\
\tableline
0&-20.2&0.68\\
500&-20.6&0.97\\
1000&-21.0&0.91\\
1500&-21.4&0.86\\
2000&-21.6&0.61\\
\end{tabular}
\tablenote{ $*$ ${\rm J_\nu\; is\;in\;units\;of\;ergs\;cm^{-2}\;s^{-1}\;
Hz^{-1}\;Sr^{-1}}$}
\end{table}
\newpage
\begin{table}
\caption{Results of proximity effect calculations taking into account the
effect of
peculiar velocities}
\begin{tabular}{cccc}
\multicolumn{1}{c}{}&\multicolumn{3}{c}{$\chi^2$
probability for}\\
\multicolumn{1}{c}{v(Km s$^{-1}$)}&\multicolumn {1}{c}{${\rm
J_\nu^*}= 10^{-20}$}&\multicolumn {1}{c}{${\rm J_\nu^*}=
10^{-20.5}$}&\multicolumn {1}{c}{${\rm J_\nu^*}= 10^{-21}$}\\
\tableline
1000&0.0000&0.0000&......\\
1500&0.0036&0.1527&......\\
2000&0.7468&0.6032&0.0009\\
2500&0.6578&0.5733&0.0096\\
3000&......&0.5044&......\\
\end{tabular}
\tablenote{ $*$ ${\rm J_\nu\; is\;in\;units\;of\;ergs\;cm^{-2}\;s^{-1}\;
Hz^{-1}\;Sr^{-1}}$}
\end{table}
\end{document}